\documentclass[a4paper]{jpconf}
\usepackage{graphicx}
\usepackage{amssymb}
\usepackage{amsmath}

\newcommand {\vect}[1]{\mbox{\boldmath $#1$}}

\newcommand {\dif}[3][]{\frac{d^{#1}#2}{d#3^{#1}}}
\newcommand {\pdif}[3][]{\frac{\partial^{#1}#2}{\partial#3^{#1}}}

\newcommand {\Alfven}{Alfv\'{e}n}
\newcommand {\Alfvenic}{Alfv\'{e}nic}
\newcommand{\sgn}{{\rm sgn}}

\begin{document}
\title{Quasi all-speed schemes for magnetohydrodynamic flows in a wide range of Mach numbers}

\author{Takashi Minoshima}
\address{Center for Mathematical Science and Advanced Technology, Japan Agency for Marine-Earth Science and Technology, 3173-25, Showa-machi, Kanazawaku, Yokohama 236-0001, Japan}
\ead{minoshim@jamstec.go.jp}

\begin{abstract}
We present novel numerical schemes for ideal magnetohydrodynamic (MHD) simulations aimed at enhancing stability against numerical shock instability and improving the accuracy of low-speed flows in multidimensions.
Stringent benchmark tests confirm that our scheme is more robust against numerical shock instability and is more accurate for low-speed, nearly incompressible flows than conventional shock-capturing schemes.
Our scheme is a promising tool for tackling MHD systems, including both high and low Mach number flows.
\end{abstract}

\section{Introduction}\label{sec:introduction}
A magnetohydrodynamic (MHD) simulation is an indispensable tool for investigating macroscopic dynamics of laboratory, space, and astrophysical plasmas.
In the field of compressible MHD simulations, a family of upwind, shock-capturing schemes have been developed based on the solution to the Riemann problem in one-dimensional hyperbolic conservation laws.
These schemes enable us to effectively resolve shocks and discontinuities in supersonic flows.
Numerous studies have been dedicated to the development of numerical schemes to ensure the robustness and accuracy of MHD simulations.
Notably, the Roe \cite{1988JCoPh..75..400B,1998ApJS..116..119B} and the Harten-Lax-van Leer discontinuities (HLLD) \cite{2005JCoPh.208..315M} approximate Riemann solvers are extensively implemented in modern MHD simulation codes.

In practical MHD simulations, however, conventional shock-capturing schemes may encounter numerical difficulties.
For high Mach number flows, these schemes tend to be vulnerable to numerical shock instability when a multidimensional shock closely aligns with the grid spacing \cite{1994IJNMF..18..555Q}.
Conversely, in the case of low Mach number flows, these schemes lead to a degradation in solution accuracy as the Mach number decreases.
In the field of computational aerodynamics, a family of the advection upstream splitting method (AUSM) \cite{1993JCoPh.107...23L,1996JCoPh.129..364L} has been extensively employed due to its high flexibility, and has been extended to the ``all-speed'' regime to permit reliable simulations in a wide range of Mach numbers, including nearly incompressible flows and high Mach number shocks \cite{2006JCoPh.214..137L,2011AIAAJ..49.1693S,2013JCoPh.245...62K}.

In this paper, we present novel numerical schemes designed for MHD simulations including both low and high Mach number flows.
The first scheme is the multistate low-dissipation advection upstream splitting method (MLAU) \cite{2020ApJS..248...12M}, an MHD extension of the all-speed AUSM scheme.
The second scheme is the low-dissipation HLLD scheme (LHLLD) \cite{2021JCoPh.110639M}, which incorporates techniques from the MLAU scheme, allowing users of the HLLD scheme to easily enjoy its capabilities.
Both schemes provide accurate solutions for nearly incompressible flows that remain super-{\Alfvenic}, a feature we refer to as {\it quasi all-speed} capability.

\section{Numerical schemes}\label{sec:numerical-schemes}
We consider one-dimensional ideal MHD equations written in the following conservative form:
\begin{eqnarray}
&& \pdif{\vect{U}}{t} + \pdif{ \vect{F}}{x} = 0,\label{eq:1}\\
&&\vect{U} = 
\begin{bmatrix}
 \rho, 
 {\rho u}, 
 {\rho v}, 
 {\rho w}, 
 {B_y}, 
 {B_z}, 
 e 
\end{bmatrix}
^T,\label{eq:2}\\
&& \vect{F} = 
\begin{bmatrix}
 \rho u \\
 \rho u^2 + P + |\vect{B}|^2/2-B_x^2 \\
 \rho v u - B_x B_y \\
 \rho w u - B_x B_z \\
 B_y u - B_x v \\
 B_z u - B_x w \\
\left(e+P+|\vect{B}|^2/2\right)u - B_x\left(\vect{u}\cdot\vect{B}\right)\\
\end{bmatrix}
,\label{eq:3}
\end{eqnarray}
where $\vect{U}$ and $\vect{F}$ are the state vector of conservative variables and the corresponding flux vector, respectively, and $\rho,\vect{u}=(u,v,w),\vect{B}=(B_x,B_y,B_z)$, and $e$ are the mass density, velocity, magnetic field, and total energy density.
The gas pressure $P$ is determined from the equation of state for the ideal gas,
\begin{eqnarray}
P = \left(\gamma-1\right)\left(e-\frac{\rho |\vect{u}|^2}{2} - \frac{|\vect{B}|^2}{2}\right),\label{eq:4} 
\end{eqnarray}
where $\gamma$ is the specific heat ratio.
The solenoidal condition of the magnetic field gives $B_x = {\rm constant}$ in one dimension.
Equation (\ref{eq:1}) is discretized on computational cells into a finite volume form as follows:
\begin{eqnarray}
\dif{\vect{U}^n_i}{t} = - \frac{\vect{\hat{F}}_{i+1/2}-\vect{\hat{F}}_{i-1/2}}{\Delta x},\;\;\; \vect{U}^n_i=\frac{1}{\Delta x}\int_{x_i-\Delta x/2}^{x_i+\Delta x/2} \vect{U}(x,t_n)dx,\label{eq:5}
\end{eqnarray}
where $\vect{\hat{F}}_{i \pm 1/2}$ is the numerical flux at the interfaces of a cell $[x_i-\Delta x/2,x_i+\Delta x/2]$.
The quality of numerical solutions is significantly influenced by the assessment of the numerical flux.

\subsection{Multistate low-dissipation advection upstream splitting method (MLAU)}\label{sec:MLAU}
Following the methodology of the AUSM-family schemes, we split the flux (equation (\ref{eq:3})) into three parts, namely, advection, pressure, and magnetic tension parts, such that
\begin{eqnarray}
\vect{F} = \rho u \vect{\Phi} + P_t \vect{N} - \vect{T},\label{eq:6}
\end{eqnarray}
where,
\begin{eqnarray}
&& \vect{\Phi}=\left(1,u,v,w,B_y/\rho,B_z/\rho,h\right)^T,\; \label{eq:7}\\
 && \vect{N}=\left(0,1,0,0,0,0,0\right)^T,\label{eq:8}\\
&&  \vect{T}= B_x\left(0,B_x/2,B_y,B_z,v,w,\vect{u}_t \cdot \vect{B_t}\right)^T,\label{eq:9}\\
&& \vect{u}_t=(0,v,w),\vect{B}_t=(0,B_y,B_z),\label{eq:78}\\
&& h=\gamma P/(\gamma-1)\rho+|\vect{u}|^2/2+|\vect{B}_t|^2/\rho,\label{eq:10}
\end{eqnarray}
and
\begin{eqnarray}
P_t=P+|\vect{B}_t|^2/2,\label{eq:11}
\end{eqnarray}
is the total pressure (excluding the contribution of $B_x$).
Subsequently, the numerical flux at the interface is
\begin{eqnarray}
&& \vect{\hat{F}}_{i+1/2}=\dot{m}\left(d_L\vect{\Phi}_{i+1/2,L}+d_R\vect{\Phi}_{i+1/2,R}\right) + \hat{P}_{t,i+1/2}\vect{N}-\vect{\hat{T}}_{i+1/2},\label{eq:15}\\
&& d_L=\frac{1+\sgn(\dot{m})}{2},d_R=\frac{1-\sgn(\dot{m})}{2},\dot{m}=\left(\rho u\right)_{i+1/2},\label{eq:16}
\end{eqnarray}
where the subscripts $L$ and $R$ refer to the left and right states at the interface. 
The mass flux $\dot{m}$ and the pressure flux $\hat{P}_t$ are built based on the all-speed AUSM scheme \cite{2006JCoPh.214..137L}, which utilizes a weighted average of the left and right states with respect the Mach number of the fast magnetosonic speed $c_f$.
The magnetic tension flux $\vect{\hat{T}}$ is built to be consistent with the HLLD scheme.

Liou argued that the scheme becomes vulnerable to numerical shock instability in multidimensions when the mass flux $\dot{m}$ includes the term proportional to the pressure difference $(P_{i+1/2,R}-P_{i+1/2,L})$ and aligns with the shock surface; although, this term is effective for stabilizing one-dimensional shocks \cite{2000JCoPh.160..623L}.
To prevent the instability while maintaining the robustness of one-dimensional shocks, we introduce a shock-detecting factor $\theta$ and multiply it by the pressure difference term (see equation (\ref{eq:23}) in the next subsection),
\begin{eqnarray}
\theta &=& {\rm min}\left(1,\frac{-{\rm min}(\Delta u,0)+{c}_{f}}{-{\rm min}(\Delta v, \Delta w, 0)+{c}_{f}}\right)^4,\label{eq:18}\\
\Delta u &=& u_{i+1,j,k}-u_{i,j,k},\label{eq:19}\\
\Delta v &=& {\rm min}(v_{i,j,k}-v_{i,j-1,k},v_{i,j+1,k}-v_{i,j,k},\nonumber \\
&& v_{i+1,j,k}-v_{i+1,j-1,k},v_{i+1,j+1,k}-v_{i+1,j,k}),\label{eq:20}\\
\Delta w &=& {\rm min}(w_{i,j,k}-w_{i,j,k-1},w_{i,j,k+1}-w_{i,j,k},\nonumber \\
&& w_{i+1,j,k}-w_{i+1,j,k-1},w_{i+1,j,k+1}-w_{i+1,j,k})\label{eq:21},
\end{eqnarray}
which quickly approaches zero when the shock surface is nearly parallel to the $z-x$ or $x-y$ plane (i.e., the shock normal direction is orthogonal to $x$).

The pressure flux $\hat{P}_t$ is expressed as a combination of the central average, pressure difference, and the velocity difference of the left and right states.
While the pressure difference term scales with the fluid velocity and acts as upwinding, the velocity difference term scales with the fast magnetosonic speed and serves as diffusion in one-dimensional compressible flows.
However, it causes excessive diffusion for low-speed flows in multidimensions.
To correct the scaling of the pressure flux, we introduce a modified fast magnetosonic speed $c_u$,
\begin{eqnarray}
c^2_{u} = \frac{1}{2}\left[\left(c^2_a + |\vect{u}|^2\right) + \sqrt{\left(c^2_a + |\vect{u}|^2\right)^2-4|\vect{u}|^2 c^2_{ax}}\right],\label{eq:22}
\end{eqnarray}
where $c_a^2=|\vect{B}|^2/\rho$ and $c_{ax}^2=B_x^2/\rho$ are {\Alfven} speeds.
We then multiples $c_u/c_f$ by the velocity difference term.
This correction appropriately scales down the pressure flux to $|\vect{u}| = c_a$, and as a result, it is expected to provide accurate solutions for super-{\Alfvenic} flows, particularly in high-beta and low-speed plasma.
For more comprehensive information about the scheme design, please refer to \cite{2020ApJS..248...12M,2021JCoPh.110639M}.

\subsection{Low-dissipation HLLD approximate Riemann solver (LHLLD)}\label{sec:LHLLD}
The novel techniques used in the MLAU scheme are implemented to the HLLD scheme to enhance stability against numerical shock instability and improve the accuracy of low-speed flows.
The normal velocity $S_M$ and the total pressure $P_{t}^*$ in the Riemann fan are modified to incorporate the shock detection in equation (\ref{eq:18}) and the pressure correction using equation (\ref{eq:22}), such that
\begin{eqnarray}
 S_M &=& \frac{(S_R-u_R)\rho_R u_R - (S_L-u_L)\rho_L u_L - \theta (P_{tR}-P_{tL})}{(S_R-u_R)\rho_R -  (S_L-u_L)\rho_L}, \label{eq:23}\\
P_{t}^* &=& \frac{(S_R-u_R)\rho_R P_{tL} - (S_L-u_L)\rho_L P_{tR}+\phi \rho_L\rho_R(S_R-u_R)(S_L-u_L)(u_R-u_L)}{(S_R-u_R)\rho_R - (S_L-u_L)\rho_L},\label{eq:24}
\end{eqnarray}
where $S_{L}={\rm min}(0,{\rm min}(u_L,u_R)-{\rm max}(c_{f,L},c_{f,R}))$ and $S_{R}={\rm max}(0,{\rm max}(u_L,u_R)+{\rm max}(c_{f,L},c_{f,R}))$.
The factor $\phi$ multiplied by the third term in equation (\ref{eq:24}) should satisfy the condition $\propto c_u/c_f \; (c_u/c_f \ll 1)$ to improve the accuracy of low Mach number flows and equal $1 \; (c_u/c_f \geq 1)$ to reduce to the original scheme at high Mach numbers.
We use
\begin{eqnarray}
 \phi = \chi (2-\chi), \;\;\; \chi={\rm min}(1,{\rm max}(c_{u,L},c_{u,R})/{\rm max}(c_{f,L},c_{f,R})).\label{eq:25}
\end{eqnarray}
For more comprehensive information about the scheme design, please refer to \cite{2021JCoPh.110639M}.

\section{Numerical experiments}\label{sec:numerical-tests}
We conduct stringent numerical experiments to assess the capability of the present schemes.
Since the numerical results obtained using the MLAU and LHLLD schemes are nearly indistinguishable, we present the LHLLD scheme for clarity.
Our numerical code incorporates the second-order MUSCL scheme with a minmod limiter for interpolating physical variables, the third-order strong stability preserving Runge-Kutta method for time integration, and the central upwind constrained transport method to preserve the solenoidal condition of the magnetic field \cite{2019ApJS..242...14M}.
We employ a CFL number of 0.4.

The first problem is the two-dimensional Kelvin-Helmholtz instability (KHI) in nearly incompressible flows.
The initial condition has a velocity shear $\vect{u}=(V_0/2)\tanh((y-10\lambda)/\lambda)\vect{e}_x$, uniform density and pressure $\rho=\rho_0,P=P_0$, and a uniform magnetic field $\vect{B}=B_0(\cos\theta,0,\sin\theta)$, with $\rho_0=V_0=B_0=\lambda=1.0$ and $\theta=71.565^{\circ}$.
We use $\gamma=2.0$.
To initiate the instability, we impose a small perturbation into the $y$-component of the velocity along the shear layer with a wavelength of $20 \lambda$, corresponding to the fastest growing mode.
The computational domain covers the range from $0 \leq x<20 \lambda$ and $0 \leq y<20 \lambda$, and it is resolved by $N \times N$ cells, where $N=64,128,256$, and $1024$ for reference.
The boundary condition is periodic and symmetric in the $x$ and $y$ directions.

\begin{figure}
\begin{center}
\includegraphics[clip,angle=0,scale=0.3]{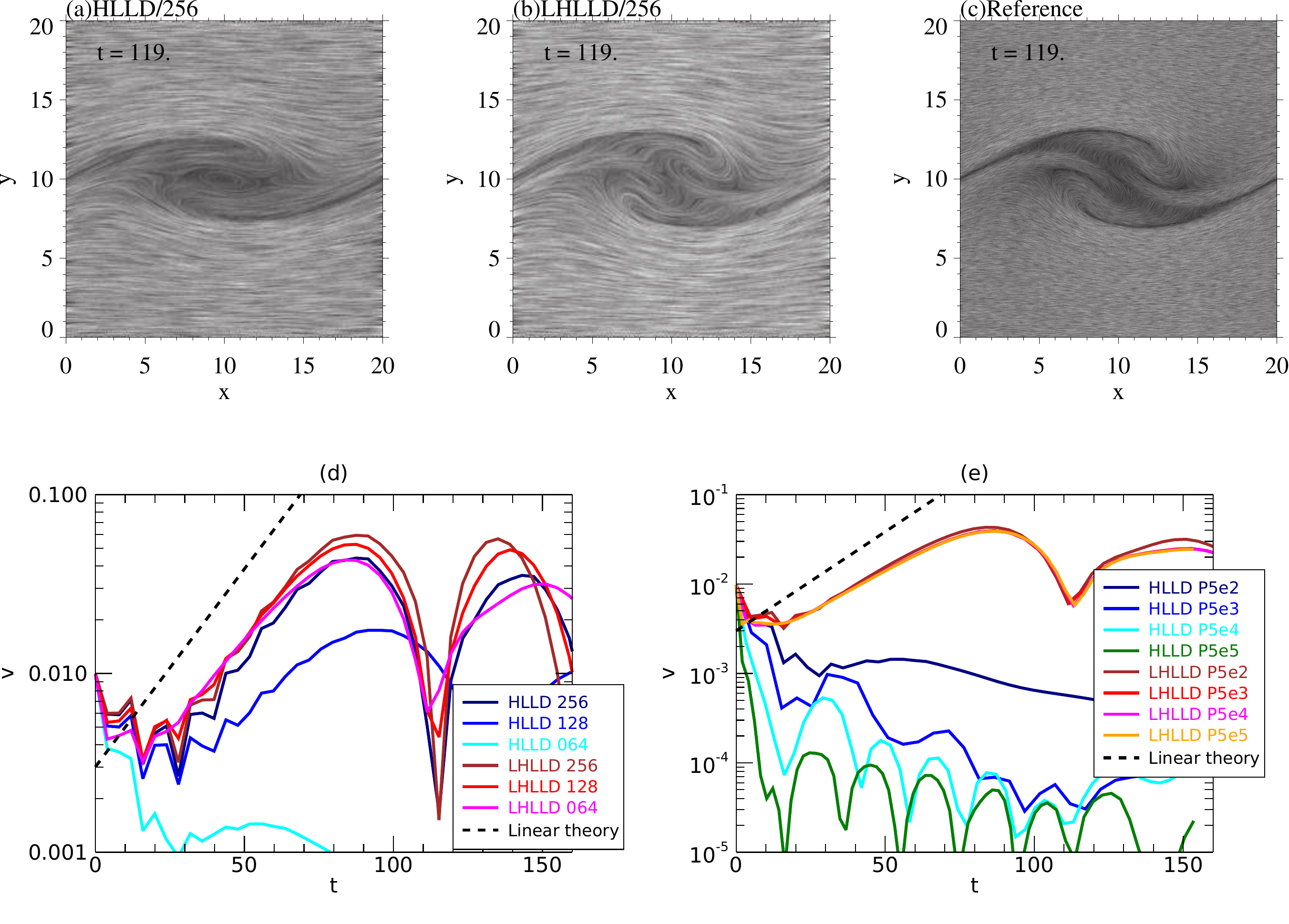}
\caption{Numerical simulation of the Kelvin-Helmholtz instability: (a-c) Stream lines obtained using the HLLD scheme at $N=256$, the LHLLD scheme at $N=256$, and the HLLD scheme at $N=1024$. (d-e) Time profiles of the fastest growing mode of the $y$-component of the velocity, demonstrating the dependence on grid resolution and the initial pressure.}
\label{fig:khi}
\end{center}
\end{figure}

Figure \ref{fig:khi}(a)-(c) shows the stream lines at $t=119$ with $P_0=500$ (corresponding Mach number of $0.0158$), obtained using the HLLD and LHLLD schemes at $N=256$, and the HLLD scheme at $N=1024$ as a reference.
The in-plane magnetic field is amplified by stretching, and the tension force distorts the flow within the vortex in $5<x<15$ and $7<y<13$.
It is evident that the LHLLD scheme outperforms the HLLD scheme in resolving the flow pattern, exhibiting good agreement with the reference solution that has $4^2$ times higher grid resolution.
Figure \ref{fig:khi}(d) shows the growth rate at $N=64,128,256$ with $P_0=500$.
The linear growth shown in a dashed line, obtained by solving the linearized MHD equations (equation (68) in \cite{2020ApJS..248...12M}), is mostly independent of the pressure in nearly incompressible flows.
The LHLLD scheme shown in red lines converges to a linear growth rate of $\gamma = 0.051V_0/\lambda$ for all runs, while the HLLD scheme shown in blue lines requires $N=256$ to achieve convergence.
 Figure \ref{fig:khi}(e) shows the growth rate at $N=64$ with $P_0=500$, $5000$, $50000$, and $500000$.
Thanks to the pressure correction (equation (\ref{eq:24})), the solutions obtained using the LHLLD scheme (red lines) are mostly independent of the pressure, while the HLLD scheme (blue lines) gets worse as the pressure increases.
Consequently, we can conclude that the LHLLD scheme possesses the capability to provide accurate solutions for nearly incompressible flows.

The second problem is the two-dimensional Richtmyer-Meshkov instability (RMI) occurring in hypersonic flows with a very weak magnetic field.
The initial condition is as follows:
\begin{eqnarray}
(\rho,\vect{u},\vect{B},P)=
\left\{
\begin{array}{c}
(1.0,0,-1.0,0,0.000034641,0,0,0.00006) \;\;\; (y>0)\\
(3.9988,0,-0.250075,0,0.000138523,0,0,0.75) \;\;\; (y<0)
\end{array}
\right.\label{eq:12}
\end{eqnarray}
which satisfies the Rankine-Hugoniot condition for perpendicular MHD shocks. 
The upstream Mach number is $100$ and the plasma beta is $\beta=2P/|\vect{B}|^2=10^5$.
We use $\gamma=5/3$.
A corrugated contact discontinuity is imposed in the upstream region, $y_{\rm cd}=Y_{0} + \psi_0 \cos (2 \pi x/\lambda)$, where $Y_{0} = 1.0$, $\psi_0=0.1$ is the corrugation amplitude, and $\lambda=1.0$ is the wavelength.
We shift a frame moving at $v=-0.625$, which corresponds to the interface velocity after the collision with the incident shock, ensuring that the structure of the RMI remains at approximately $y=0$ throughout the simulation run.
The computational domain covers the range from $0 \leq x < \lambda$ and $-40 \lambda \leq y < 40 \lambda$, and it is resolved by $N \times 80N$ cells, where $N=128$.
The boundary condition is periodic in the $x$ direction and is fixed to the initial state in the $y$ direction.
\begin{figure}
\begin{center}
\includegraphics[clip,angle=0,scale=0.22]{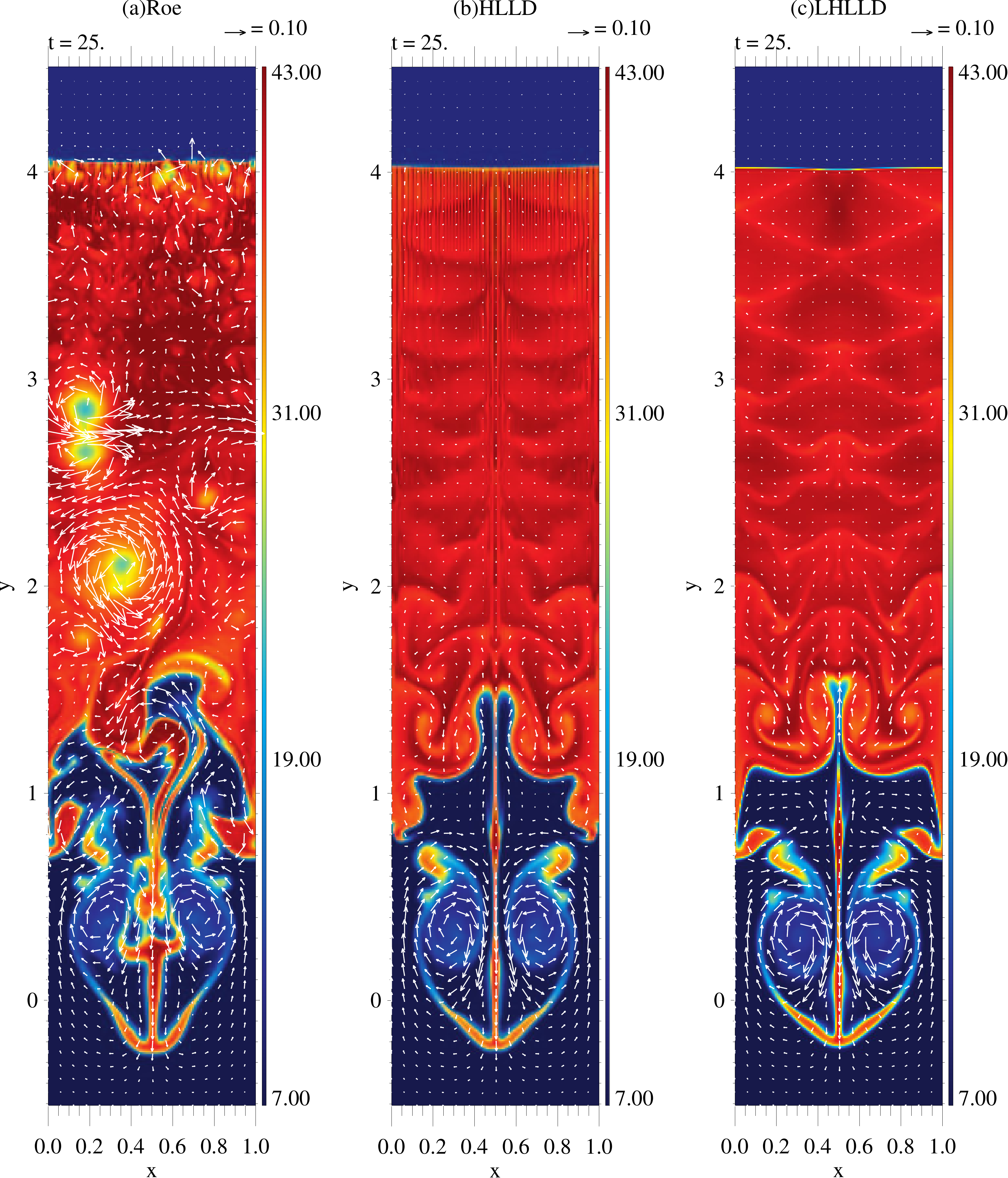}
\caption{Density profiles in the Richtmyer-Meshkov instability at $t=25$, obtained using (a) the Roe scheme, (b) the HLLD scheme, and (c) the LHLLD scheme. The arrows represent the rotational velocity component.}
\label{fig:rmi}
\end{center}
\end{figure}

Figure \ref{fig:rmi} shows the density profile at $t=25$ obtained using the Roe, HLLD, and LHLLD schemes.
The surface at $y=4$ corresponds to the transmitted shock, and the distinctive mushroom-shaped structure in $0<y<1$ corresponds to the nonlinear evolution of the RMI.
In panel (a), the solution obtained using the Roe scheme is severely distorted by numerical shock instability.
The instability generates multiple unphysical voids in the shocked region, disrupts the structure of the region of interest $(0<y<1)$, and causes a loss of solution symmetry.
In panel (b), the HLLD scheme significantly mitigates the catastrophic structures observed in the Roe scheme, but it still exhibits grid-scale oscillation around the shock front.
In contrast, the LHLLD scheme in panel (c) succeeds in completely eliminating unphysical structures by utilizing the shock detection, resulting in a smooth solution.
Furthermore, the solution obtained using the LHLLD scheme exhibits faster roll-up flows in the region $0<y<1$ compared to the HLLD scheme, which is attributed to the pressure correction for subsonic flows in the Mach number range of $0.01-0.1$.

The last problem is the three-dimensional MHD turbulence induced by the magnetorotational instability (MRI).
Building upon our previous research \cite{2015ApJ...808..54M}, we solve isothermal MHD equations including the Coriolis force within the framework of a local shearing box approximation \cite{1995ApJ...440..742H}.
We apply the orbital advection method to update the velocity fluctuation $\vect{u'}=\vect{u}-\vect{u}_0$, where $\vect{u}_0=-q\Omega x \vect{e}_y$ corresponds to the background shear flow (with $q=1.5$ and $\Omega=1$ representing angular frequency) \cite{2010ApJS..189..142S} .
Since the velocity fluctuation typically remains subsonic, the solution may be affected by the pressure correction in the present scheme.
The initial condition is uniform, $(\rho,\vect{u'},\vect{B},P)=(1,0,B_0 \vect{e}_z,P_0)$, with an incompressible uniform random noise imposed to $u'$ to initiate the instability.
The ambient magnetic field $B_0$ is fixed to $0.025$, and the gas pressure $P_0$ is set to $3.125,312.5,3125$, yielding initial plasma beta values of $10^4,10^6,10^7$.
The computational domain covers the range from $-1<x<1,-2<y<2$, and $-0.5<z<0.5$, and it is resolved by $64\times256\times32$ cells.
The wavelength of the fastest growing mode of the MRI, $\lambda_{\rm FGM}=2 \pi B_0 / \sqrt{\rho} \Omega=0.16$, is resolved by approximately five grid points.
The boundary condition is periodic in the $y$ and $z$ directions, and shearing periodic in the $x$ direction.
\begin{figure}
\begin{center}
\includegraphics[clip,angle=0,scale=0.33]{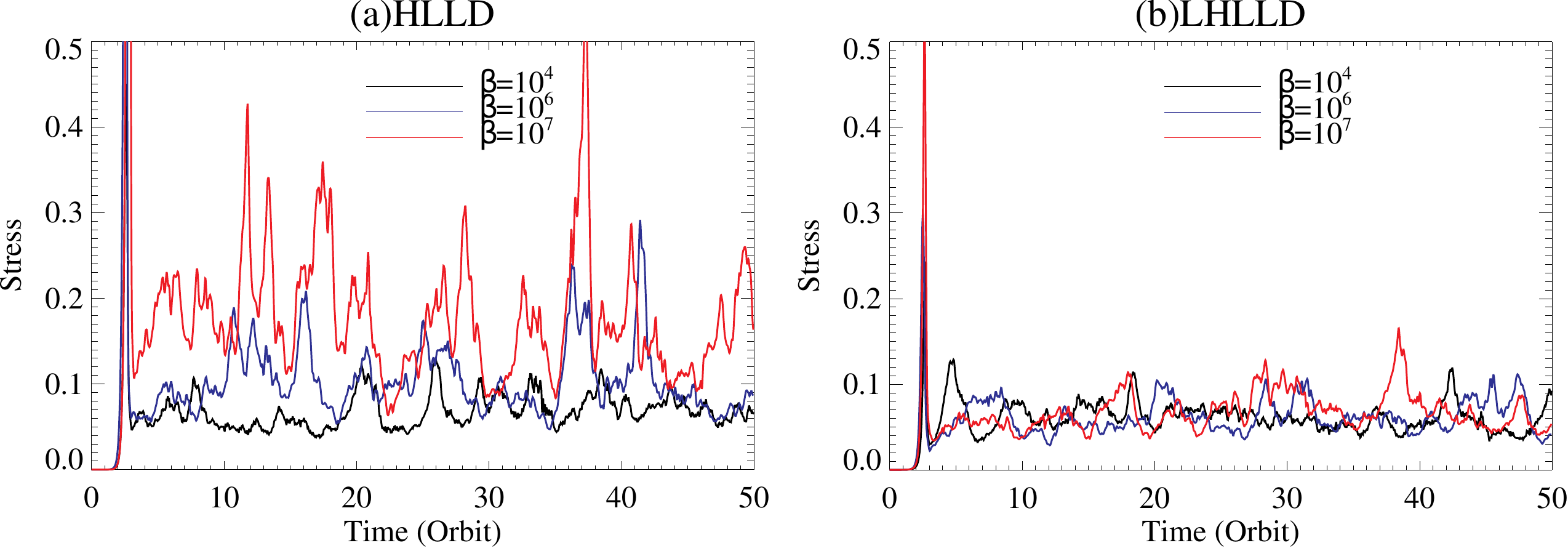}
\caption{Time profiles of the volume-averaged stress in the MRI-induced turbulence with plasma beta values of $10^4$ (black lines), $10^6$ (blue lines), and $10^7$ (red lines), obtained using (a) the HLLD scheme and (b) the LHLLD scheme.}
\label{fig:mri}
\end{center}
\end{figure}

\begin{table}
\caption{Time- and volume-averaged quantities in the MRI-induced turbulence.}\label{tab:mritable}
\begin{center}
\begin{tabular}{lllllll}
\br
Scheme & $P_0$ & $\beta$ & $w$ & $\Delta P/P_0$ & $\vect{\lambda}$  for $u'$ & $\vect{\lambda}$ for $B_y$\\
\mr
HLLD& $3.125 \times 10^{0}$ & $10^4$&$7.0\times 10^{-2}$ &$5.8\times 10^{-2}$ & $(1.0,1.6,0.64)$& $(0.55,1.3,0.49)$\\
HLLD& $3.125 \times 10^{2}$ & $10^6$&$1.1\times 10^{-1}$ &$7.0\times 10^{-4}$ & $(1.5,2.0,0.66)$& $(0.55,1.3,0.48)$\\
HLLD& $3.125 \times 10^{3}$ & $10^7$&$1.8\times 10^{-1}$ &$1.2\times 10^{-4}$ & $(1.8,2.3,0.68)$& $(0.62,1.5,0.50)$\\
\mr
LHLLD& $3.125 \times 10^{0}$ &$10^4$&$6.3\times 10^{-2}$ &$5.8\times 10^{-2}$ & $(1.0,1.6,0.62)$& $(0.54,1.3,0.49)$\\
LHLLD& $3.125 \times 10^{2}$ &$10^6$&$6.1\times 10^{-2}$ &$4.4\times 10^{-4}$ & $(1.0,1.4,0.58)$& $(0.52,1.2,0.48)$\\
LHLLD& $3.125 \times 10^{3}$ &$10^7$&$6.6\times 10^{-2}$ &$4.8\times 10^{-5}$ & $(1.0,1.4,0.57)$& $(0.54,1.3,0.49)$\\
\br
\end{tabular} 
\end{center} 
\end{table}

Figure \ref{fig:mri} shows the time profiles of the volume-averaged stress $w=\rho u'v' - B_x B_y/4\pi$ obtained using the HLLD and LHLLD schemes.
Notably, these results exhibit discrepancies that become more pronounced as the gas pressure (plasma beta) increases.
The time- and volume-averaged quantities over the time interval $t=25-50$ are summarized in Table \ref{tab:mritable}.
For various pressure values, the stress values remain approximately constant around $0.065$ with the LHLLD scheme, while they increase from $0.07$ to $0.18$ with the HLLD scheme.
Considering that the flow remains subsonic within the Mach number range of $0.005-0.2$, we can conclude that the pressure-independent solution provided by the LHLLD scheme is indicative of incompressible turbulence.
This conclusion is supported by examining the time- and volume-averaged pressure perturbation $\Delta P=P-P_0$, presented in the fifth column of Table \ref{tab:mritable}.
The LHLLD scheme satisfies the asymptotic behavior in the incompressible limit $\Delta P/P_0=O(M^2)=O(P_0^{-1})$.
\begin{figure}
\begin{center}
\includegraphics[clip,angle=0,scale=0.25]{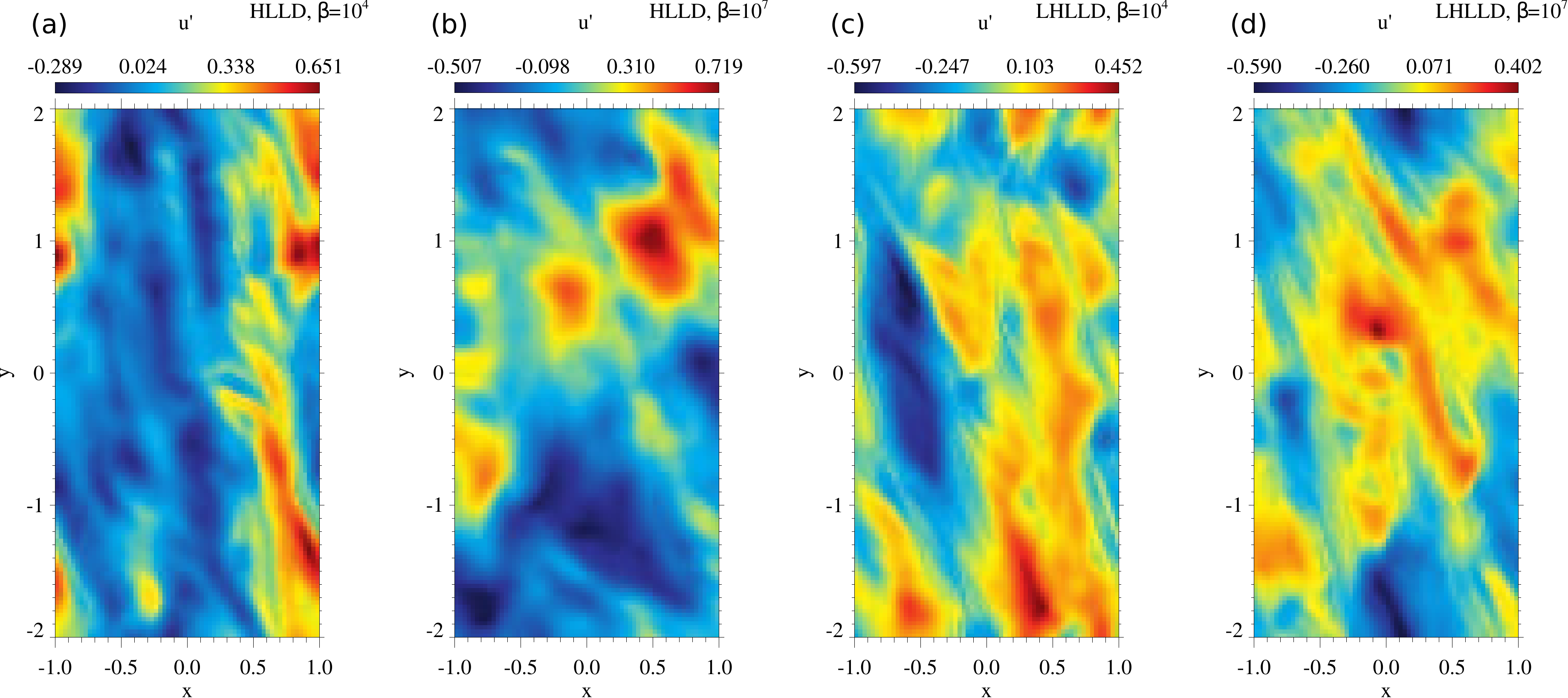}
\caption{Snapshots of $u'(x,y,z=0,t=50)$ in the MRI-induced turbulence with plasma beta values of $10^4,10^7$, obtained using (a,b) the HLLD scheme and (c,d) the LHLLD scheme.}
\label{fig:mri2d}
\end{center}
\end{figure}

The discrepancy in the solutions between the HLLD and LHLLD schemes can be attributed to the dependence of the saturation level of the MRI-induced turbulence on the magnetic Prandtl number \cite{2007MNRAS.378.1471L}.
Specifically, ideal MHD simulations of the MRI-induced turbulence are sensitive to the numerical magnetic Prandtl number, which is determined by the scheme design and is not necessarily specified \cite{2015ApJ...808..54M}.
The numerical magnetic Prandtl number is defined as the ratio of the numerical viscosity $\nu_n$ to resistivity $\eta_n$, both of which arise from the numerical diffusion terms involved in the momentum and induction equations.
For subsonic flows, the numerical viscosity is evaluated from the pressure in the Riemann fan (equation (\ref{eq:24})), resulting in $\nu_{n,{\rm HLLD}}=c_f \Delta x$ and $\nu_{n,{\rm LHLLD}}=c_u \Delta x$.
As shown in Figure \ref{fig:mri2d}, the velocity profile reveals that small-scale structures become less pronounced in the HLLD scheme as the gas pressure increases, while the structure in the LHLLD scheme remains similar regardless of the pressure.
Meanwhile, the numerical resistivity is identical in both schemes, and is evaluated from an explicit expression of the magnetic tension term, yielding $\eta_{n,{\rm HLLD}}=\eta_{n,{\rm LHLLD}}={\rm max}\left(|\vect{u}|,c_{ax}\right) \Delta x$ \cite{2020ApJS..248...12M}.
Consequently, it is anticipated that the numerical magnetic Prandtl number in the HLLD scheme increases with increasing pressure, while it remains approximately constant in the LHLLD scheme.

As a proxy for the numerical diffusion coefficients, we measure characteristic wavelengths defined as follows:
\begin{eqnarray}
 (\lambda_x,\lambda_y,\lambda_z) = \frac{2 \pi}{(k_x,k_y,k_z)},
\;\;\; (k_x,k_y,k_z)=\sqrt{\frac{1}{\overline{f^2}} \left(\overline{\left(\pdif{f}{x}\right)^2}, \overline{\left(\pdif{f}{y}\right)^2}, \overline{\left(\pdif{f}{z}\right)^2}\right)},\label{eq:13}
\end{eqnarray}
where $\overline{f}$ represents the volume average of the variable $f$.
The time-averaged characteristic wavelengths of $u'$ and $B_y$ are presented in the sixth and seventh columns of Table \ref{tab:mritable}.
The wavelength of $u'$ in the HLLD scheme tends to increase as the gas pressure increases, while other quantities remain approximately constant.
This suggests that the numerical viscosity and the numerical magnetic Prandtl number in the HLLD scheme are likely to increase with increasing pressure, causing the increase in the saturation level \cite{2015ApJ...808..54M}; in contrast, these values remain constant in the LHLLD scheme, resulting in the pressure-independent solution.

\section{Conclusion}\label{sec:conclusion}
We presented novel numerical schemes designed for MHD flows in a wide range of Mach numbers.
The schemes, the multistate low-dissipation advection upstream splitting method (MLAU) \cite{2020ApJS..248...12M} and the low-dissipation HLLD approximate Riemann solver (LHLLD) \cite{2021JCoPh.110639M}, enhance stability against numerical shock instability and improve the accuracy of low-speed flows in multidimensions by incorporating shock detection and pressure correction.
Stringent benchmark tests confirm the performance of the scheme, outperforming conventional shock-capturing schemes.
The scheme provides accurate solutions for nearly incompressible flows that remain super-{\Alfvenic}, by virtue of the pressure correction that scales down appropriately to $|\vect{u}|=c_a$.
We refer to this capability as {\it quasi all-speed} for super-{\Alfvenic} flows.
Importantly, this feature ensures that the numerical magnetic Prandtl number remains approximately constant, a critical condition for the numerical study of MHD turbulent phenomena.
With this novel ability, the scheme becomes a promising tool to tackle MHD systems, including both high and low Mach number flows.

The source code written in C++ programming language is available on GitHub\footnote{{https://github.com/minoshim/qasMHD}}.

\ack{We would like to thank T. Miyoshi, K. Kitamura, S. Hirose, and T. Sano for their valuable contributions and insightful discussions.
 This work is supported in part by JSPS KAKENHI Grant Number JP20K04056.}

\section*{References}
\bibliographystyle{iopart-num}
\bibliography{ref}

\end{document}